\documentclass[twocolumn,superscriptaddress,amsmath, amssymb, amsfonts,preprintnumbers,aps,prd,longbibliography,nofootinbib]{revtex4-1}

\usepackage{scalerel}
\usepackage{xcolor}
\usepackage{amsmath,amsfonts,amssymb}
\usepackage[small,bf,hang]{caption}
\usepackage{slashed}
\usepackage{latexsym,epsfig}
\usepackage{dsfont}
\usepackage{arydshln}
\usepackage{extarrows}
\usepackage{hyperref}
\usepackage{array}

\def\moth{\mathsurround=0pt}
\newdimen\zo \zo=0pt

\def\tick{\leaders\hrule height 0.5ex depth 0pt \hskip 0.5pt}
\def\upboxfill{$\moth \setbox\zo\hbox{\tick}%
  \hskip 3pt\hbox to 0pt{$\tick$\hss}\hrulefill \hbox to 7.5pt{$\tick$\hss}$}

\def\dtick{\leaders\hrule height .34pt depth 0.5ex \hskip 0.5pt}
\def\downboxfill{$\moth \setbox\zo\hbox{\dtick}%
  \hskip 2pt\hbox to 0pt{$\dtick$\hss}\hrulefill \hbox to 2pt{$\dtick$\hss}$}

\def\bec{\begin{center}}
\def\ec{\end{center}}

\def\nn{\nonumber}

\def\be{\begin{equation}}
\def\ee{\end{equation}}
\def\bea{\begin{eqnarray}}
\def\eea{\end{eqnarray}}
\def\ba{\begin{array}}
\def\ea{\end{array}}

\begin{document}

\title{Constructing Conformal Double Field Theory through a Double Copy Map}

\author{Eric Lescano} 
\email{eric.lescano@uwr.edu.pl}
\affiliation{Institute for Theoretical Physics (IFT), University of Wroclaw, \\
pl. Maxa Borna 9, 50-204 Wroclaw,
Poland}

\author{Jesús A. Rodríguez}
\email{jarodriguez@df.uba.ar}
\affiliation{Universidad de Buenos Aires, FCEyN, Departamento de Física, Ciudad Universitaria, 1428 Buenos Aires, Argentina}


\begin{abstract}
We follow the classical Double Copy (DC) procedure that links Yang-Mills and Double Field Theory (DFT), and we apply it on a four-derivative gauge theory which is known to be related to Weyl gravity at the level of the amplitudes. We obtain a perturbative T-duality invariant theory on a double geometry, or Conformal Double Field Theory (CDFT), incorporating Weyl gravity plus $b$-field and dilaton contributions at quadratic order, without the need to impose a gauge fixing condition. We also extend the formulation to cubic order for the case of vanishing generalized dilaton, which still incorporates Weyl gravity when $\Box h_{\mu \nu}=h=0$. CDFT, together with ordinary DFT, are examples of T-duality invariant theories constructed through classical DC maps, revealing a promising and deep connection between gauge theories and T-duality invariant models.  
   
\end{abstract}

\maketitle

\section{Introduction}

The quadratic and cubic contributions to perturbative Double Field Theory (DFT) \cite{DFT1,DFT2} can be constructed by performing an off-shell classical Double Copy (DC) map \cite{Dcopy1,Dcopy2,Dcopy3,Dcopy4,Dcopy5,class1,class2,class3,class4,class5,class6,class7} to the (quadratic and cubic) Yang-Mills Lagrangian \cite{HohmDC}. Subsequent studies have delved deeper into the algebraic structures underlying this connection up to quartic order \cite{HohmDC21,HohmDC22,HohmDC23}, providing a strong foundation for constructing quartic DFT as the DC of the quartic contributions coming from the Yang-Mills Lagrangian. These powerful results open up the possibility of extending this method to other gauge theories, leading to new T-duality invariant formulations on the double geometry, beyond DFT. In this context, DFT would be just one particular example of a family of classical T-duality invariant double copies, revealing a deep connection between gauge theories and T-duality invariant theories.

A recent proposal that explores this possibility was presented in \cite{LMR}, where the authors analyzed the higher-derivative gauge model outlined in \cite{DFJ1,DFJ2,DFF1,DFF2,DFF3}. The DC of this model results in a double gravitational theory defined in terms of a generalized vielbein and a generalized dilaton. This theory is related to quadratic Weyl gravity when the supergravity fields satisfy $g_{\mu\nu}=h_{\mu\nu}$ and $b_{\mu \nu}=\varphi=0$ (pure gravitational case), and the gauge fixing condition $\Box h - \partial_{\mu}\partial_{\nu}h^{\mu \nu}=0$ is imposed. The results in \cite{LMR} suggest that constructing a non-perturbative Conformal Double Field Theory (CDFT) is possible, while it can be formulated independently of specific gauge choices.

Remarkably, the reverse engineering of constructing gauge theories from double models through a zeroth and a single copy in the double geometry is possible thanks to the pioneering work of \cite{KL}, along with further advances in this direction \cite{KSextra1,KSextra2,KSextra3,KSextra4,KSextra5,KSextra6}. While ordinary DFT has proven to be a powerful framework for studying the classical DC program, the results of this work show the existence of a CDFT, whose single and zeroth copy should be related to the formulation in \cite{DFJ1,DFJ2,DFF1,DFF2,DFF3}. In other words, our model paves the way towards constructing a non-perturbative CDFT, offering a promising approach to constructing T-duality invariant theories as the classical DC of a given gauge theory.

In this work, we improve upon the model in \cite{LMR} by including the dynamics of a charged scalar field $\phi_{\alpha}$. The DC contributions from this field are crucial for introducing the missing quadratic terms, which enable us to construct the quadratic CDFT without requiring any gauge fixing condition. In this way, we demonstrate that it is possible to construct a CDFT through an off-shell and classical DC map, which reduces to quadratic Weyl gravity (plus $b$-field and dilaton contributions) after imposing the strong constraint. We also go beyond the quadratic order of the CDFT. Although developing the full cubic theory is a challenging endeavor, we focus on the scenario with a vanishing generalized dilaton. In this simplified context, we construct the cubic Lagrangian by further extending the DC map, fixing all coefficients by comparing the pure gravitational limit, where $\Box h_{\mu \nu}=h=0$ (stemming from the vanishing generalized dilaton condition), to cubic Weyl gravity under the same conditions. This approach uniquely determines all coefficients in this reduced scenario.

The key findings of this work are as follows: The model presented in \cite{LMR} is enhanced by incorporating a charged scalar field, essential for deriving the complete quadratic CDFT Lagrangian without gauge fixing. The model is further extended to cubic order. By leveraging the gauge theory from \cite{DFJ1,DFJ2} and applying the DC map, the cubic CDFT Lagrangian with a vanishing generalized dilaton is constructed. All undetermined coefficients in this Lagrangian are fixed by comparison with cubic Weyl gravity under the constraint $\Box h_{\mu \nu}=h=0$.

\section{From Gauge Theories to T-Duality Invariant Models: State of the Art}

In \cite{HohmDC}, the authors derived a classical and off-shell DC map to construct the quadratic and cubic perturbative formulations of Double Field Theory (DFT)\footnote{For reviews, see \cite{DFTreview1,DFTreview2,DFTreview3} and the second lecture of \cite{Electure}.} from the Yang-Mills Lagrangian. This process can be summarized schematically as follows:
\bea
\left[\textrm{YM}(A_{\mu}{}^{a}) \xrightarrow{DC} \textrm{DFT}(e_{\mu \bar \nu},\Phi) \xrightarrow{SC} \textrm{SuGRA}\right]^{(2,3)}\, , \nn
\eea
where (2,3) represents the quadratic and cubic contributions of these theories. SuGRA refers to the NS-NS sector of supergravity, which is obtained from DFT after applying the strong constraint (SC). The notation (DC) stands for the DC procedure, which we will explain in detail below.

The method involves expressing the quadratic (cubic) contributions of Yang-Mills theory in momentum space. Following this, the gauge fields, metrics, and structure constants are identified with gravitational objects in a double geometry by promoting the color indices of the gauge group to a second set of gravitational indices. For instance, we have the identifications
\bea
\label{idHohm1}
A_{\mu}{}^{a}(k) & \longrightarrow & e_{\mu\bar{\nu}}(k,\bar{k}) \, , \\
\kappa_{ab} & \longrightarrow & \frac{1}{2}(\eta^{\bar{\mu}\bar{\nu}}-\frac{\bar{k}^{\bar{\mu}}\bar{k}^{\bar{\nu}}}{k^2}) \, ,
\label{idHohm2}
\eea
for the gauge field and the Cartan-Killing metric. The notation for the indices in the diagram, and in the rest of the paper, is as follows
\begin{itemize}
\item $a,b,\dots$: gauge color indices (adj. rep.),
\item $\mu, \nu,\dots$: space-time indices, 
\item $\bar{\mu},\bar{\nu},\dots$: space-time indices.
\end{itemize} 
The dependence of the fields is now doubled to include the \textit{dual} momenta $\bar{k}$ in the same footing as the regular momenta $k$, given rise to a double geometric structure such as in DFT. Returning to configuration space, the resulting theory has the exact form of DFT. More precisely, at quadratic order it has the structure of the gauge invariant double field theory, while at cubic order it is the cubic double field theory action subject to a gauge condition that originates from Siegel gauge in string field theory \cite{HohmDC}.

With this strategy in mind, the authors in \cite{LMR} presented a procedure to introduce higher-derivative corrections in a way that the obtained double theory is related to conformal symmetry. The starting point is the higher-derivative gauge Lagrangian
\be
{\cal L} = \frac{1}{2}\kappa_{ab}D_{\mu}F^{\mu\nu a}D_{\rho}F^{\rho}{}_{\nu}{}^{b} \, ,
\label{DFDFold}
\ee
on which the authors apply the same procedure as in \cite{HohmDC}. The DC map of the gauge Lagrangian (\ref{DFDFold}) gives rise to a higher order, double geometrical theory
\be
\left[\textrm{DFDF}(A_{\mu}{}^{a}) \xrightarrow{DC} \textrm{CDFT}(e_{\mu\bar{\nu}},\Phi)\right]^{(2,3)}\, \nn ,
\ee
depending on the fundamental fields of perturbative DFT. This theory further reduce to quadratic Weyl gravity in the already mentioned pure gravitational case\footnote{In general, when the Kalb-Ramond field and the dilaton are non-vanishing, matter contributions must be considered.
} 
(pg: $g_{\mu\nu}=h_{\mu\nu}$ and $b_{\mu \nu}=\varphi=0$),
\be
\left[\textrm{CDFT}(e_{\mu\bar{\nu}},\Phi) \xrightarrow{pg} \textrm{gauge-fixed Weyl gravity}(h_{\mu \nu})\right]^{(2)}\, \nn ,
\ee
subject to the gauge-fixing condition
\be
\Box h - \partial_{\mu}\partial_{\nu}h^{\mu\nu}=0 \, .
\label{gaugefixingold}
\ee

The necessity of imposing the previous gauge-fixing condition implies that it is not possible to generate all the terms in the quadratic Weyl Lagrangian. This limitation highlights a significant problem: the lack of a gauge-independent formulation of the theory restricts our ability to construct a fully non-perturbative version of the model.

In order to address this limitation, our approach enhances the model in \cite{LMR} by including a charged scalar field, $\phi_{\alpha}$. The gauge model that we propose is given by
\bea
{\cal L} & = & a_{1}\kappa_{ab}D_{\mu}F^{\mu\nu a}D_{\rho}F^{\rho}{}_{\nu}{}^{b} + a_{2}\kappa^{\alpha\beta}D_{\mu}\phi_{\alpha}D^{\mu}\phi_{\beta}\,  \nn \\ 
& & + a_{3}f_{abc}F_{\mu}{}^{\nu a}F_{\nu}{}^{\lambda b}F_{\lambda}{}^{\mu c} + a_{4}C^{\alpha}{}_{ab}\phi_{\alpha}F_{\mu\nu}{}^{a} F^{\mu\nu b}\, \nn \\
& & + a_{5}d^{\alpha\beta\gamma}\phi_{\alpha}\phi_{\beta}\phi_{\gamma}\, ,
\label{fullgauge_sect_2}
\eea
where the $a_{i}$ are real coefficients to be determined. The scalar field transforms under the action of the same gauge group but in the real representation, denoted by color indices $\alpha,\beta,\dots$, and it is allowed to interact with the gauge field $A_{\mu}{}^{a}$. This model is inspired in the construction \cite{DFJ1,DFJ2}, and it is related to the DC of Weyl gravity at the level of scattering amplitudes.

As we are interested in constructing a double theory that includes conformal gravity in the appropriate limit, our strategy for determining the coefficients in the action will be to expand the Lagrangian \eqref{fullgauge_sect_2} to quadratic and cubic order and to compare the reduction in a convenient limit, the pure gravity limit, with/without generalized dilaton contributions for the quadratic/cubic case, respectively. In the cubic case, we will also consider the condition $\Box h_{\mu \nu}=h=0$, which is fully compatible with the vanishing of the generalized dilaton. The next two sections are devoted to present this construction.

\section{Quadratic Conformal Double Field Theory}

Starting from the theory proposed in \eqref{fullgauge_sect_2}, the only source of quadratic contributions comes from the kinetic terms of the two fundamental fields. Specifically, we have:
\be
{\cal L}^{(2)} = 4a_{1}\kappa_{ab}\partial_{\mu}\partial^{[\mu}A^{\nu]a}\partial^{\rho}\partial_{[\rho}A_{\nu]}{}^{b} + a_{2}\kappa^{\alpha\beta}\partial_{\mu}\phi_{\alpha}\partial^{\mu}\phi_{\beta}\, .
\ee
Following the method outlined in \cite{HohmDC,LMR}, we integrate by parts and then transform this action into momentum space, yielding:
\bea
S_{\mathrm{DC}}^{(2)} & = & - \int d^{D}k\left[a_{1}k^{4}\kappa_{ab}\Pi^{\mu\nu}(k)A_{\mu}{}^{a}(k)A_{\nu}^{b}(-k)\right.\, \nn \\ 
& & \ \ \ \ \ \ \ \ \ \ \ \ \ \left. + a_{2}k^{2}\kappa^{\alpha\beta}\phi_{\alpha}(k)\phi_{\beta}(-k)\right]\, .
\eea
Where $\Pi^{\mu\nu}(k) = \eta^{\mu\nu} - \frac{k^{\mu}k^{\nu}}{k^{2}}$, suggests the identification \eqref{idHohm2} for the Cartan-Killing metric, which complements the identification \eqref{idHohm1} for the gauge field. At this stage, it is necessary to identify the objects associated with the kinetic term of the scalar field to complete the DC prescription for our model. This leads to
\bea
\label{id1}
\phi_{\alpha}(k) & \longrightarrow & k_{\mu}e^{\mu\bar{\nu}}(k,\bar{k}) + 2\bar{k}^{\bar{\nu}}\Phi(k,\bar{k})\, , \\
\kappa^{\alpha \beta} & \longrightarrow & \frac{\bar{k}_{\bar{\mu}}\bar{k}_{\bar{\nu}}}{k^{2}}\, ,
\label{id2}
\eea
where $\Phi(k,\bar{k})$ is the DFT generalized dilaton in momentum space. The identification of the scalar field is strongly indicated by the missing quadratic terms noted in \cite{LMR}, which are essential to completing the quadratic contributions of conformal gravity without relying on a specific gauge.

The quadratic DC Lagrangian, after performing all the identifications, is given by
\bea
\label{DFDF_DFT}
S_{\rm{DC}}^{(2)} & = & - \frac{1}{2}\int d^{D}x d^{D}\bar{x}\left[a_{1}\left(\Box e^{\mu\bar{\nu}}\Box e_{\mu\bar{\nu}} - \Box e^{\mu\bar{\nu}}\partial_{\mu}\partial^{\rho}e_{\rho\bar{\nu}}\right.\right.\, \nn \\ 
& & - \left. \Box e^{\mu\bar{\nu}}\bar{\partial}_{\bar{\nu}}\bar{\partial}^{\bar{\sigma}}e_{\mu\bar{\sigma}} + \partial^{\mu}\bar{\partial}^{\bar{\nu}}e_{\mu\bar{\nu}}\partial^{\rho}\bar{\partial}^{\bar{\sigma}}e_{\rho\bar{\sigma}}\right)\, \nn \\ 
& & \left. - 2a_{2}\left(\partial_{\mu}\partial_{\bar{\nu}}e^{\mu\bar{\nu}} + 2\Box\Phi\right)^{2}\right]\, .
\eea
This action reproduces the action obtained in \cite{LMR} but includes new contributions coming from the DC of the scalar field $\phi_{\alpha}$.

Now we consider the pure gravity limit, which demands $x=\bar{x}$ (in this context this is equivalent to solve the strong constraint in DFT), together with $b_{\mu \nu}=\varphi=0$. The previous action becomes
\bea
S_{\rm{DC}}^{(2)} & = & - \frac{1}{2}\int d^{D}x\Big[a_{1}\left(\Box h^{\mu\nu}\Box h_{\mu\nu} - 2\Box h^{\mu\nu}\partial_{\mu}\partial^{\rho}h_{\rho\nu} \right. \Big. \nn \\  
& & \left. \left. + \ \partial^{\mu}\partial^{\nu}h_{\mu\nu}\partial^{\rho}\partial^{\lambda}h_{\rho\lambda}\right) - 2a_{2}\left(\Box h - \partial_{\mu}\partial_{\nu}h^{\mu\nu}\right)^{2}\right]\, \nn ,
\eea
which present the same structure that the quadratic contributions of the Weyl gravity action (see Appendix A). In fact, if we demand both actions to be the same, the coefficients $a_1,a_2$ are fixed as 
\bea
a_1 & = & -2\left(\frac{D-3}{D-2}\right) \, , \\
a_2 & = & -\frac{1}{(D-1)}\left(\frac{D-3}{D-2}\right)\, .
\label{coeffquadratic}
\eea

It is notable, though expected, that the action vanishes in $D=3$, as in three dimensions, the Weyl tensor is identically zero. The terms associated with the coefficient $a_{1}$ in \eqref{DFDF_DFT} were already obtained in \cite{LMR}. The new result presented here is the emergence of terms with a non-zero coefficient $a_{2}$, which arise from the scalar field dynamics. Although these terms were absent in \cite{LMR}, the authors in that work compensated for this absence through the gauge-fixing condition (\ref{gaugefixingold}) in the pure gravity limit.

\section{Cubic Conformal Double Field Theory with $\Phi=0$}

\subsection{Constructing the cubic double copy}

Once the quadratic contributions of our model have been obtained and the coefficients $a_{1}$ and $a_{2}$ have been calculated, ensuring that, in the pure gravity limit, the model reduces to Weyl gravity, we are ready to proceed with the calculation of the cubic theory. Now, we will use the full Lagrangian introduced in \eqref{fullgauge_sect_2}
\bea
{\cal L} & = & a_{1}\kappa_{ab}D_{\mu}F^{\mu\nu a}D_{\rho}F^{\rho}{}_{\nu}{}^{b} + a_{2}\kappa^{\alpha\beta}D_{\mu}\phi_{\alpha}D^{\mu}\phi_{\beta}\,  \nn \\ 
& & + a_{3}f_{abc}F_{\mu}{}^{\nu a}F_{\nu}{}^{\lambda b}F_{\lambda}{}^{\mu c} + a_{4}C^{\alpha}{}_{ab}\phi_{\alpha}F_{\mu\nu}{}^{a} F^{\mu\nu b}\, \nn \\
& & + a_{5}d^{\alpha\beta\gamma}\phi_{\alpha}\phi_{\beta}\phi_{\gamma}\, .
\eea
In this Lagrangian, $f_{abc}$ are the structure constants in the adjoint representation of the gauge group, $C_{\alpha ab}$ are the Clebsch-Gordan coefficients, and $d_{\alpha \beta \gamma}$ is a totally symmetric object defined through the generators $(T_{R}^{a})^{\alpha \beta}$ of the real representation. The gauge covariant derivatives are defined as
\bea
D_{\rho}F_{\mu\nu}{}^{a} & = & \partial_{\rho}F_{\mu\nu}{}^{a} + gf^{a}{}_{bc}A_{\rho}{}^{b}F_{\mu\nu}{}^{c}\, , \nn \\
D_{\mu}\phi^{\alpha} & = & \partial_{\mu}\phi^{\alpha} - ig (T_{R}^{a})^{\alpha \beta} A_{\mu a} \phi_{\beta}\, \nn ,
\eea
with the Yang-Mills curvature defined in the standard way $F_{\mu\nu}{}^{a} = 2\partial_{[\mu}A_{\nu]}{}^{a} + gf^{a}{}_{bc}A_{\mu}{}^{b}A_{\nu}{}^{c}$.

Our goal in this section is to construct the gravitational cubic contributions arising from the DC prescription of this model, determining the remaining coefficients $a_3$, $a_4$, and $a_5$ in the same way as we did for the quadratic terms. While analyzing the full Lagrangian is a very ambitious task, we will focus on the case with vanishing generalized dilaton \cite{HohmDC}
\be
\Phi(k,\bar{k}) = \frac{1}{k^{2}}k^{\mu}\bar{k}_{\bar{\nu}}e_{\mu}{}^{\bar{\nu}} = 0 \, .
\ee
 
To construct the cubic Lagrangian of CDFT , we have to identify the following quantities
\bea
\label{idf}
f_{abc} & \longrightarrow & i \frac{a_6}{a_1} (\eta^{\mu\nu}k_{12}^{\rho} + \eta^{\nu\rho}k_{23}^{\mu} + \eta^{\rho\mu}k_{31}^{\nu})  \,  \\
\label{idC}
C_{\alpha ab} & \longrightarrow &  i k^{\bar \mu} \eta^{\bar \nu \bar \rho} \, ,\\
\label{idomega}
(Tr^{a})^{\alpha \delta} A^{\sigma}{}_{a} & \longrightarrow & i \Gamma^{\sigma}_{\bar \mu \bar \nu} \, , \\
\label{idd}
d_{\alpha \beta \gamma} & \longrightarrow &  \frac{i}{k^2} k^{\bar \mu} k^{\bar \nu} k^{\bar \rho}\, .
\eea
In the above identifications, the only object requiring an additional coefficient to fix its cubic contributions is $f_{abc}$. In this case, the coefficient $a_1$ is inherited from the cubic contribution of the $DFDF$ term, which needs to be canceled.  

We observe that the Weyl tensor is non-trivial in $D>3$ and therefore $a_1^{-1}$ is well defined. The identification (\ref{idf}) follows the same structure as in \cite{HohmDC},but the inclusion of the $\frac{a_1}{a_6}$ coefficients introduces a global factor in this identification, which is crucial for reproducing Weyl gravity in arbitrary dimensions, as we will demonstrate in the next part of this section. The identification (\ref{idomega}) illustrates a natural correspondence between a gauge connection and a gravitational connection for the generalized vielbein.

The cubic DC Lagrangian with $\Phi=0$ is given by
\begin{widetext}
\bea
\label{Cubic_HD_DFT}
S_{\rm{DC}}^{(3)}|_{\Phi=0} & = &\int d^{D}x \ d^{D}\bar{x} (-\frac{a_6}{2}) \left[\Box\partial^{\rho}\bar{\partial}^{\bar{\sigma}}e_{\mu\bar{\nu}}e^{\mu\bar{\nu}}e_{\rho\bar{\sigma}} - \Box\bar{\partial}^{\bar{\sigma}}e_{\mu\bar{\nu}}\partial^{\rho}e^{\mu\bar{\nu}}e_{\rho\bar{\sigma}} + \Box\bar{\partial}^{\bar{\sigma}}e_{\mu\bar{\nu}}\partial^{\mu}e^{\rho\bar{\nu}}e_{\rho\bar{\sigma}} - \Box\bar{\partial}^{\bar{\sigma}}e_{\mu\bar{\nu}}e^{\rho\bar{\nu}}\partial^{\mu}e_{\rho\bar{\sigma}}\right.\, \nn \\
& & \ \ \ \ \ \ \ \ \ \ \ \ - \Box e_{\mu\bar{\nu}}\partial^{\mu}\bar{\partial}^{\bar{\sigma}}e^{\rho\bar{\nu}}e_{\rho\bar{\sigma}} + \Box e_{\mu\bar{\nu}}\bar{\partial}^{\bar{\sigma}}e^{\rho\bar{\nu}}\partial^{\mu}e_{\rho\bar{\sigma}} - \partial^{\mu}\partial^{\rho}\partial^{\lambda}\bar{\partial}^{\bar{\sigma}}e_{\mu\bar{\nu}}e_{\rho}{}^{\bar{\nu}}e_{\lambda\bar{\sigma}} + \partial^{\mu}\partial^{\rho}\bar{\partial}^{\bar{\sigma}}e_{\mu\bar{\nu}}\partial^{\lambda}e_{\rho}{}^{\bar{\nu}}e_{\lambda\bar{\sigma}}\, \nn \\
& & \ \ \ \ \ \ \ \ \ \ \ \ - \partial^{\rho}\partial^{\lambda}\bar{\partial}^{\bar{\sigma}}e_{\mu\bar{\nu}}\partial^{\mu}e_{\rho}{}^{\bar{\nu}}e_{\lambda\bar{\sigma}} + \partial^{\rho}\partial^{\lambda}\bar{\partial}^{\bar{\sigma}}e_{\mu\bar{\nu}}e_{\rho}{}^{\bar{\nu}}\partial^{\mu}e_{\lambda\bar{\sigma}} + \partial^{\rho}\partial^{\lambda}e_{\mu\bar{\nu}}\partial^{\mu}\bar{\partial}^{\bar{\sigma}}e_{\rho}{}^{\bar{\nu}}e_{\lambda\bar{\sigma}} - \partial^{\rho}\partial^{\lambda}e_{\mu\bar{\nu}}\bar{\partial}^{\bar{\sigma}}e_{\rho}{}^{\bar{\nu}}\partial^{\mu}e_{\lambda\bar{\sigma}}\, \nn \\
& & \ \ \ \ \ \ \ \ \ \ \ \ \left. - \bar{\partial}^{\bar{\nu}}\bar{\partial}^{\bar{\sigma}}\bar{\partial}^{\bar{\kappa}}\partial^{\rho}e_{\mu\bar{\nu}}e^{\mu}{}_{\bar{\sigma}}e_{\rho\bar{\kappa}} + \bar{\partial}^{\bar{\nu}}\bar{\partial}^{\bar{\sigma}}\bar{\partial}^{\bar{\kappa}}e_{\mu\bar{\nu}}\partial^{\rho}e^{\mu}{}_{\bar{\sigma}}e_{\rho\bar{\kappa}} + \bar{\partial}^{\bar{\nu}}\bar{\partial}^{\bar{\sigma}}e_{\mu\bar{\nu}}\bar{\partial}^{\bar{\kappa}}\partial^{\mu}e^{\rho}{}_{\bar{\sigma}}e_{\rho\bar{\kappa}} - \bar{\partial}^{\bar{\nu}}\bar{\partial}^{\bar{\sigma}}e_{\mu\bar{\nu}}\bar{\partial}^{\bar{\kappa}}e^{\rho}{}_{\bar{\sigma}}\partial^{\mu}e_{\rho\bar{\kappa}}\, \right]\, \nn \\ &&
\ \ \ \ \ \ \ \ \ \ \ \ + a_{2} \partial_{\sigma}\partial_{\lambda} e^{\lambda \bar \epsilon} \Gamma^{(1)\sigma}{}_{\bar \epsilon \bar \tau} \partial_{\epsilon} e^{\epsilon \bar \tau} + 3 a_3 \left[\partial_{\mu} e^{\nu \bar \rho} \partial_{\bar \rho}\partial_{\nu}e^{\lambda}{}_{\bar \epsilon} \partial_{\lambda}e^{\mu \bar \epsilon} - \partial_{\mu} e^{\nu \bar \rho} \partial_{\nu}e^{\lambda}{}_{\bar \epsilon} \partial_{\bar \rho}\partial_{\lambda}e^{\mu \bar \epsilon} \right]  \nn \\ &&
\ \ \ \ \ \ \ \ \ \ \ \ - 3 a_{3} \left[\partial_{\mu} \partial_{\bar \epsilon} e^{\nu \bar \rho} \partial_{\nu}e^{\lambda}{}_{\bar \rho} \partial^{\mu}e_{\lambda}{}^{\bar \epsilon} - \partial_{\mu} e^{\nu \bar \rho} \partial_{\bar \epsilon} \partial_{\nu}e^{\lambda}{}_{\bar \rho} \partial^{\mu}e_{\lambda}{}^{\bar \epsilon}\right] - 3 a_{3} \left[\partial_{\mu} e^{\nu \bar \rho} \partial_{\nu}\partial_{\bar \rho}e^{\lambda \bar \sigma} \partial^{\mu}e_{\lambda \bar \sigma} - \partial_{\mu} e^{\nu \bar \rho} \partial_{\nu}e^{\lambda \bar \sigma} \partial^{\mu}\partial_{\bar \rho}e_{\lambda \bar \sigma}\right]  
\nn \\ && \ \ \ \ \ \ \ \ \ \ \ \ 
- 3 a_{3} \left[\partial_{\mu} e^{\nu \bar \rho} \partial_{\nu}e^{\lambda \bar \epsilon} \partial^{\mu}\partial_{\bar \epsilon}e_{\lambda \bar \rho} - \partial_{\mu} \partial_{\bar \epsilon} e^{\nu \bar \rho} \partial_{\nu}e^{\lambda \bar \epsilon} \partial^{\mu}e_{\lambda \bar \rho}\right] + a_{4} \partial_{\bar \mu}\partial_{\epsilon} e^{\epsilon \bar \mu} \partial_{[\sigma} e_{\lambda]}{}^{\bar \nu} \partial^{\lambda}e^{\sigma}{}_{\bar \nu} \, ,
\label{fullLagrangian}
\eea    
\end{widetext}
where the notation $\Gamma^{(1)}{}^{\rho}_{\bar \mu \bar \nu}$ refers to the first order perturbation of the connection. It is possible to see that the Lagrangian \eqref{fullgauge_sect_2} produces terms proportional to the generalized dilaton associated to the coefficient $a_{5}$, and therefore this coefficient does not appear in this Lagrangian. To determine the remaining coefficients, we will compare (\ref{fullLagrangian}) with the cubic contributions of Weyl gravity. 

We will perform this analysis under the assumption of a vanishing wave equation $\Box h_{\mu \nu}=0$ and the traceless condition $h=0$. Given these conditions, the generalized dilaton will not contribute to the gravitational terms after parametrization. Furthermore, our method cannot fix the coefficient $a_5$, as its contributions depend solely on the generalized dilaton, which vanishes in this case.

In the next section, we will demonstrate that this method is sufficiently robust to uniquely determine the coefficients $a_3$, $a_4$ and $a_6$.

\subsection{Fixing the coefficients when $\Phi=0$}

The cubic contributions for Weyl gravity are given by
\bea
S_{\rm{CG}}^{(3)} & = & \int d^{D}x \left[\frac{1}{2}hC^{(1)}_{\mu\nu\rho\lambda}C^{(1)\mu\nu\rho\lambda} - 4h^{\mu\nu}C^{(1)}_{\mu\rho\lambda\sigma}C^{(1)}_{\nu}{}^{\rho\lambda\sigma} \right. \nn \\ && \left. + 2C^{(2)}_{\mu\nu\rho\lambda}C^{(1)\mu\nu\rho\lambda}\right]\, .
\eea
Imposing the conditions  
\be
\Box h_{\mu \nu}=h=0 \, ,
\label{conditions}
\ee
which are fully compatible with $\Phi=0$, and performing a series of integrations by parts with the help of CADABRA \cite{Cadabra}, the cubic Lagrangian can be reduced to the following two contributions,
\bea
L^{(3)} & = &  f(D)\partial_{\delta}\partial_{\lambda}h^{\epsilon \lambda}\partial^{\nu}h_{\nu \sigma}\partial_{\epsilon}h^{\sigma \delta}\, , \nn \\ 
& & + g(D)\partial_{\mu}\partial_{\sigma}h^{\mu}{}_{\epsilon} \partial_{\delta}\partial_{\lambda}h^{\sigma \epsilon}h^{\delta \lambda}\, ,
\eea
where 
\bea
f(D) & = &\frac{-2(D-3)(11D+15)}{5(D-2)(D-1)} \, , \nn \\
g(D) & = & \frac{(D-3)(34D+5)}{5(D-2)(D-1)} \, . 
\eea
As before, we find that $D=3$ is the only physically meaningful root of both $f(D)$ and $g(D)$, consistent with the physical interpretation of the Weyl tensor. When the double conformal model (\ref{fullLagrangian}) is parametrized and restricted to the pure gravitational limit, it takes the following form

\bea
L^{(3)}_{DC} = && -(\frac{a_6}{4} + 3 a_3) \partial_{\delta} \partial_{\lambda}h^{\epsilon \lambda} \partial^{\nu}h_{\nu \sigma} \partial_{\epsilon}h^{\sigma \delta} \nn \\ && + \frac{15}{4} a_6 \partial_{\mu} \partial_{ \sigma}h^{\mu}{}_{ \epsilon} \partial_{\delta}\partial_{\lambda}{h^{\sigma \epsilon}} h^{\delta \lambda} \nn \\ &&+(-a_6+ \frac{a_4}{2}) \partial_{\mu}\partial_{ \epsilon}h^{\mu \epsilon}\partial_{\sigma}h_{\lambda \nu} \partial^{\nu}h^{\sigma \lambda}) \, .
\eea
Therefore, the coefficients are uniquely fixed as
\bea
a_3= - \frac{1}{45} g(D) - \frac{1}{3}f(D) \, , \\
a_4 = 2 a_{6} = \frac{8}{15} g(D) \, \, .  
\label{cubicfix}
\eea
The preceding identifications determine the double action (\ref{fullLagrangian}), which represents the full cubic CDFT with a vanishing generalized dilaton. In \eqref{cubicfix}, we observe the necessity of the $a_{6}$ coefficient in identifying the structure constants. Additionally, we note that both the $F^{3}$ term and the interaction term between the gauge field and the scalar field are essential for reproducing Weyl gravity in this limit.

By comparing with \cite{LMR}, it becomes evident that both the charged scalar field and the cubic self-interactions of the gauge field $A_{\mu}{}^{a}$, are necessary to accurately match Weyl gravity in the pure gravitational scenario.

\section{Conclusions}

We construct the quadratic Lagrangian for a T-duality invariant theory which is the classical double copy of the Lagrangian \eqref{fullgauge_sect_2}. The field theory content of this theory is exactly the same as in DFT with a generalized frame and a generalized dilaton. After parametrization and imposing the pure gravitational solution, the theory reduces to quadratic Weyl gravity. We refer to this theory as Conformal Double Field Theory (CDFT).

The model presented in this paper enhances the formulation provided in \cite{LMR} by incorporating the dynamics of a charged scalar field, following a proposal outlined in \cite{DFJ1,DFJ2}. The inclusion of the charged scalar field proves to be crucial for constructing both the quadratic and cubic contributions of the CDFT. Notably, the quadratic contributions from the scalar field enable us to reproduce Weyl gravity plus matter contributions after imposing the strong constraint, without the necessity of gauge fixing. The global coefficients in the initial gauge Lagrangian must be fixed using \eqref{coeffquadratic} to accurately embed Weyl gravity.

In this work, we also construct the cubic structure of the CDFT with a vanishing generalized dilaton, and determine its coefficients by imposing the conditions (\ref{conditions}) and comparing it to cubic Weyl gravity. The only coefficient that remains undetermined when analyzing the full cubic theory is $a_5$, as it depends exclusively on the generalized dilaton. It would be interesting to fix the remaining coefficient by including the generalized dilaton and relaxing the condition $h=0$ in the pure gravity limit. Alternatively, one could consider constructing a non-perturbative CDFT directly using T-duality multiplets. While the generalized Riemann tensor of DFT is not fully determined, there might be a clever approach to fixing the coefficients in the bi-parametric higher-derivative DFT constructed in \cite{CarmenDiego} to reduce to Weyl gravity in the pure gravitational limit. The results of this work suggest a promising and profound connection between gauge theories and T-duality invariant theories through double copy maps.

\subsection*{Acknowledgements}
We are very grateful to G.Menezes and F.Hassler for discussions on the initial stage of this project. E.L. is supported by the SONATA BIS grant 2021/42/E/ST2/00304 from the National Science Centre (NCN), Poland. The work of J.A.Rodriguez is supported by CONICET.

\begin{widetext}

\appendix

\section{Quadratic Weyl Gravity}

The Weyl gravity action is given by 
\be
S_{\rm{CG}} = \int d^{D}x \ \sqrt{-g} \ C_{\mu\nu\rho\lambda}C^{\mu\nu\rho\lambda}\, ,
\label{Weyl}
\ee
where the Weyl tensor is defined as
\be
C_{\mu\nu\rho\lambda} = R_{\mu\nu\rho\lambda} - \frac{2}{D-2}\left(g_{\mu[\rho}R_{\lambda]\nu} - g_{\nu[\rho}R_{\lambda]\mu}\right) + \frac{2}{\left(D-1\right)\left(D-2\right)}Rg_{\mu[\rho}g_{\lambda]\nu}\, .
\ee
We will consider perturbations around the flat space such that 
\be 
g_{\mu\nu} = \eta_{\mu\nu} + h_{\mu\nu}\, , \qquad g^{\mu\nu} = \eta^{\mu\nu} - h^{\mu\nu} + h^{\mu\rho}h_{\rho}{}^{\nu} + \cdots\, , \qquad \sqrt{-g} = 1 + \frac{1}{2}h + \cdots\, , \qquad h=h_{\mu}{}^{\mu}\, \nn .
\ee

The quadratic contributions that comes from this set up are
\be
\label{quadratic_weyl}
S_{\rm{CG}}^{(2)} = \frac{D-3}{D-2}\int d^{D}x\left[\left(\Box h^{\mu\nu}\Box h_{\mu\nu} - 2\Box h^{\mu\nu}\partial_{\mu}\partial^{\rho}h_{\rho\nu} + \partial^{\mu}\partial^{\nu}h_{\mu\nu}\partial^{\rho}\partial^{\lambda}h_{\rho\lambda}\right) - \frac{1}{D-1}\left(\Box h - \partial_{\mu}\partial_{\nu}h^{\mu\nu}\right)^{2}\right]\, .
\ee

\end{widetext}


\begin{thebibliography}{99}
\label{SEC:Refs}
\bibitem{DFT1}
 C.~Hull and B.~Zwiebach, ``Double Field Theory," JHEP {\bf 09} (2009) 099, [hep-th/0904.4664].

\bibitem{DFT2}
  O.~Hohm, C.~Hull and B.~Zwiebach,
  ``Generalized metric formulation of Double Field Theory,"
  JHEP {\bf 08} (2010) 008,
  [hep-th/1006.4823].
  
\bibitem{Dcopy1} 

Z. Bern, J. J. M. Carrasco and H. Johansson, “New Relations for Gauge-Theory Amplitudes”,
Phys. Rev. D78 (2008) 085011 [0805.3993].

\bibitem{Dcopy2} 
Z. Bern, J. J. M. Carrasco and H. Johansson, “Perturbative Quantum Gravity as a Double Copy of Gauge Theory”, Phys. Rev. Lett. 105 (2010) 061602
[1004.0476]

\bibitem{Dcopy3} 
Z. Bern, J. J. Carrasco, M. Chiodaroli, H. Johansson, and R. Roiban, “The Duality Between Color and Kinematics and its Applications,” arXiv:1909.01358 [hep-th].

\bibitem{Dcopy4} 
T. Adamo, J. J. M. Carrasco, M. Carrillo-Gonzalez, M. Chiodaroli, H. Elvang, H. Johansson, D. O’Connell, R. Roiban and O. Schlotterer, “Snowmass White Paper: the Double
Copy and its Applications,” [arXiv:2204.06547]

\bibitem{Dcopy5} 
H. Kawai, D. C. Lewellen and S. H. H. Tye, “A Relation Between Tree Amplitudes of Closed and Open Strings”, Nucl. Phys. B269 (1986) 1.

\bibitem{class1}
R. Monteiro, D. O’Connell, and C. D. White, “Black holes and the double copy,” JHEP 12 (2014) 056, arXiv:1410.0239 [hep-th].

\bibitem{class2}
A. Luna, R. Monteiro, I. Nicholson, A. Ochirov, D. O’Connell, N. Westerberg, and C. D. White, “Perturbative spacetimes from Yang-Mills theory,” JHEP 04 (2017) 069, arXiv:1611.07508 [hep-th].

\bibitem{class3}
W. D. Goldberger and A. K. Ridgway, “Bound states and the classical double copy,” Phys. Rev. D 97 no. 8, (2018) 085019, arXiv:1711.09493 [hep-th].

\bibitem{class4}
J. Plefka, J. Steinhoff, and W. Wormsbecher, “Effective action of dilaton gravity as the classical double copy of Yang-Mills theory,” Phys. Rev. D 99 no. 2, (2019) 024021, arXiv:1807.09859 [hep-th].

\bibitem{class5}
N. E. J. Bjerrum-Bohr, P. H. Damgaard, G. Festuccia, L. Planté, and P. Vanhove, “General Relativity from Scattering Amplitudes,” Phys. Rev. Lett. 121 no. 17, (2018) 171601, arXiv:1806.04920 [hep-th].

 \bibitem{class6}
A. Brandhuber, G. Chen, G. Travaglini, and C. Wen, “Classical gravitational scattering from a gauge-invariant double copy,” JHEP 10 (2021) 118, arXiv:2108.04216 [hep-th].

 \bibitem{class7}
A. Brandhuber, G. Chen, H. Johansson, G. Travaglini, and C. Wen, “Kinematic Hopf Algebra for Bern-Carrasco-Johansson Numerators in Heavy-Mass Effective Field Theory and Yang-Mills Theory,” Phys. Rev. Lett. 128 no. 12, (2022) 121601, arXiv:2111.15649 [hep-th].

\bibitem{HohmDC}
F.~Diaz-Jaramillo, O.~Hohm and J.~Plefka,
``Double field theory as the double copy of Yang-Mills theory,''
Phys. Rev. D \textbf{105}, no.4, 045012 (2022), arXiv:2109.01153 [hep-th].

\bibitem{HohmDC21}
R. Bonezzi, F. Diaz-Jaramillo, O. Hohm, ``The gauge structure of double field theory follows from Yang-Mills theory'', Phys.Rev.D 106 (2022) 2, 026004, [hep-th/2203.07397].

\bibitem{HohmDC22}
R. Bonezzi, C. Chiaffrino, F. Diaz-Jaramillo, O. Hohm, ``Gauge invariant double copy of Yang-Mills theory: the quartic theory", arXiv: 2212.04513 [hep-th].

\bibitem{HohmDC23}
R. Bonezzi, C. Chiaffrino, F. Diaz-Jaramillo, and O. Hohm, “Weakly Constrained Double
Field Theory as the Double Copy of Yang-Mills Theory,” arXiv:2309.03289 [hep-th]

\bibitem{LMR}
E. Lescano, G. Menezes, and J. A. Rodriguez, “Aspects of Conformal Gravity and
Double Field Theory from a Double Copy Map”, Phys.Rev.D 108 (2023) 12, 126017, arXiv:2307.14538 [hep-th] 

\bibitem{DFJ1}
H. Johansson and J. Nohle, “Conformal gravity from gauge theory,” [arXiv:1707.02965] (2017).

\bibitem{DFJ2}
H. Johansson, G. Mogull and F. Teng, “Unraveling conformal gravity amplitudes,” JHEP 09, 080 (2018).

\bibitem{DFF1}
T. Azevedo, R. L. Jusinskas and M. Lize, “Bosonic sectorized strings and the (DF)$^2$ theory,” JHEP 01, 082 (2020).

 \bibitem{DFF2}
G. Menezes, “Color-kinematics duality, double copy and the unitarity method for higher-derivative QCD and quadratic
gravity,” JHEP 03, 074 (2022).

\bibitem{DFF3}
G. Menezes, “Leading Singularities in Higher-Derivative Yang–Mills Theory and Quadratic Gravity”, Universe 8 (2022) 6, 326, [hep-th/2205.04996]

\bibitem{KL}
K. Lee, ``Kerr-Schild Double Field Theory and Classical Double Copy'', JHEP \textbf{1810}(2018) 027, [hep-th/1807.08443]. 

\bibitem{KSextra1}
W. Cho, K. Lee, ``Heterotic Kerr-Schild Double Field Theory and Classical Double Copy'', JHEP \textbf{07}(2019) 030,[hep-th/1904.11650]

\bibitem{KSextra2}
E. Lescano and A. Rodr\'iguez, ``${\cal N} = 1$ Supersymmetric Double Field Theory and the
generalized Kerr-Schild Ansatz'', JHEP \textbf{10} (2020) 148, [hep-th/2002.07751].

\bibitem{KSextra3}
E. Lescano and A. Rodr\'iguez,
``Higher-derivative Heterotic Double Field Theory and Classical Double Copy'', JHEP {\bf 07} (2021) 072, [hep-th/2101.03376].

\bibitem{KSextra4}
S. Angus, K. Cho, K. Lee, ``The Classical Double Copy for Half-Maximal Supergravities and T-duality'', JHEP {\bf 10} (2021) 211 [hep-th/2105.12857]

\bibitem{KSextra5}
K. Kim, K. Lee, R. Monteiro, I. Nicholson, and D. P. Veiga, ``The Classical Double
Copy of a Point Charge'', JHEP {\bf 02} (2020) 046, [hep-th/1912.02177].

\bibitem{KSextra6}
E. Lescano, S. Roychowdhury, ``Heterotic Kerr-Schild Double Field Theory and its double Yang-Mills formulation," JHEP 04 (2022) 090, e-Print: 2201.09364 [hep-th].

\bibitem{DFTreview1}
G.~Aldazabal, D.~Marques and C.~Nu\~nez,
  ``Double Field Theory: A Pedagogical Review," Class.\ Quantum\ Grav.\  {\bf 30} (2013) 163001, [hep-th/1305.1907].

\bibitem{DFTreview2}
   O.~Hohm, D.~Lust and B.~Zwiebach,
  ``The Spacetime of Double Field Theory: Review, Remarks, and Outlook,"
  Fortsch.\ Phys.\  {\bf 61} (2013) 926,  [hep-th/1309.2977].

\bibitem{DFTreview3}
   D.~S.~Berman and D.~C.~Thompson,
  ``Duality Symmetric String and M-Theory," Phys.\ Rept.\  {\bf 566} (2014) 1, [hep-th/1306.2643].

\bibitem{Electure}
E. Lescano,``$\alpha'$--corrections and their double formulation," [hep-th/2108.12246]. 

\bibitem{Cadabra} K.~Peeters, ``Introducing Cadabra: A Symbolic computer algebra system for field theory problems,''
  hep-th/0701238.

\bibitem{CarmenDiego}
``T-duality and $\alpha'$-corrections,''
  JHEP {\bf 1510}, 084 (2015), [hep-th/1507.00652].

\end{thebibliography}
\end{document}